\documentclass[aps,preprint,preprintnumbers,nofootinbib,showpacs]{revtex4}

\usepackage{amsmath}
\usepackage{amssymb}
\usepackage{mathtools}
\usepackage{dsfont}
\usepackage{color}
\usepackage{graphicx,accents}
\usepackage[normalem]{ulem}

\def\bea{\begin{eqnarray}}
\def\eea{\end{eqnarray}}
\def\sea{\nonumber \\&&}

\def\rra{\right\rangle}

\def\ssc{\scriptscriptstyle}
\def\lsim{\mathrel{\raise.3ex\hbox{$<$\kern-.75em\lower1ex\hbox{$\sim$}}} }
\def\gsim{\mathrel{\raise.3ex\hbox{$>$\kern-.75em\lower1ex\hbox{$\sim$}}} }

\begin{document}
\draft
\preprint{{\vbox{\hbox{NCU-HEP-k095}
\hbox{May 2022}
\hbox{rev. Sep 2022}
\hbox{ed. Apr 2023}
}}}
\vspace*{.4in}


\title{\boldmath  Quantum Origin of (Newtonian) Mass and Galilean Relativity Symmetry
\vspace*{.2in}}

\author{Otto C. W. Kong}

\affiliation{Department of Physics and Center for High Energy and High Field Physics,
National Central University, Chung-li, Taiwan 32054  \\
}


\begin{abstract}
\vspace*{.1in}
The Galilei group has been taken as the fundamental symmetry 
for `nonrelativistic'  physics,  quantum or classical. Our fully 
group theoretical formulation approach to the quantum theory 
asks for some adjustments. We present a sketch of the full picture
here, emphasizing aspects that are different from the more familiar 
picture. The analysis involves a more careful treatment of the 
relation between the exact mathematics and its physical application
in the dynamical theories, and a more serious
full implementation of the mathematical logic than what is usually
available in the physics literature. The article summarizes our earlier 
presented formulation while focusing on the part beyond, with an 
adjusted, or corrected, identification of the basic representations 
having the (Newtonian) mass as a Casimir invariant and the notion
of center of mass as dictated by the symmetry, that is particularly 
also to be seen as the Heisenberg-Weyl symmetry inside it. Another 
result is the necessary exclusion of the time translational symmetry. 
The time translational symmetry in the Galilei group plays no role
in the formulation of the dynamical theory and does not correspond
to the physical time in any nontrivial setting.
\\[.2in]

\end{abstract}

\maketitle

\section{Introduction}
Symmetry consideration is one of the most important theoretical
guiding principles in modern physics. A group theoretical formulation
of fundamental theories can go a very long way in `explaining' the
fundamental structures.  Group theory, especially
in the commutator (Lie) algebra of the quantum operators which 
serves as generators of unitary symmetry (group) transformations
of the fundamental (relativity) symmetry, has been common in the 
literature of quantum physics. Yet, even most physicists today 
may have limited appreciation of what is the strict mathematical
logic and what may be only contingent on the application of that in 
a particular physical theory. For example, in the physics literature, 
the commutator algebra and the way the corresponding operators have 
been introduced are often taken as the defining structure of the 
symmetry Lie algebra. That actually is nothing more 
than one consistent representation of the symmetry. In fact, it 
is only such a representation when the Hilbert space the operators 
are supposed to act on is indeed the representation space.  The
word `representation' has been used in a few different ways,
which is also a source of confusion. In this article, we use the
word only in the straight symmetry theoretical sense as in the
literature of mathematics.

We have results that indeed go against some of the very first things
physicists learned in the early days of their education. To appreciate
them, readers certainly need to exercise patience and care to
consider our logic critically but otherwise objectively and seriously 
with an open mind. 

Relativity symmetries have been taken as the most important 
kind of fundamental symmetries in physics. The Galilean symmetry 
for `nonrelativistic' physics has been firstly taken as the symmetry 
for the Newtonian space-time which is  a representation/coset space 
of the Galilei group $G(3)$. The phase space for a Newtonian 
particle can also be identified with a $G(3)$ coset space \cite{086}. 
The Hilbert space for the states for a quantum particle is known as 
the representation space of the Heisenberg-Weyl symmetry $H(3)$,
as usually appreciated through the commutator algebra of the 
position and momentum operators. $G(3)$ symmetry is still taken
to be the relativity symmetry, though it can only be realized so-called
projectively. In fact, what is called a projective representation of 
a Lie group can, and in our opinion should only, be seen as a unitary 
representation of its $U(1)$ central extension \cite{AI}. The $U(1)$ 
central extension, $\tilde{G}(3)$, as an abstract symmetry contains 
the $H(3)$ as a subgroup. Taking that seriously gives a better 
picture of the symmetry theoretical formulation of quantum 
mechanics \cite{070}.  This article gives a slightly modified picture 
of that,  with two mathematically minor adjustments having,
however, important implications. One is to take out the time 
translation symmetry from ${G}(3)$ or $\tilde{G}(3)$. The other
is to have the central charge generator in $H(3)$ represented 
as the Newtonian mass operator. The key new features are: 
Newtonian particle mass is a Casimir invariant of the 
fundamental/relativity symmetry for the quantum theory only to be 
inherited by classical mechanics; which then dictates the exact 
notion of center of mass; the Newtonian time in the dynamical theory
does not come from the  `nonrelativistic' relativity symmetry.\footnote
{Our study here is restricted to the basic picture of particle dynamics,
classical and quantum. Otherwise, for example in relation to 
geometric description of gravity, somewhat extended symmetry
framework may be of interest \cite{D1,D2,D3}.}

The phase space for a quantum particle, is, like all classical phase 
space, a symplectic manifold \cite{gqm}. The Poisson bracket in 
terms of the observables is essentially given by the commutator
 \cite{AS,078}. Note that symplectic manifold and its Poisson 
bracket are generally very closely connected to the symmetry 
of Lie group/algebra and its Lie bracket \cite{st}. With the 
symplectic geometry, one has a picture of one-parameter groups 
of Hamiltonian flows as symmetry transformations on the phase 
space, each for a generic Hamiltonian function \cite{S}. The 
Hamiltonian functions which match to (Hermitian) elements of 
the observable algebra \cite{CMP} have the transformations as 
unitary transformations on the Hilbert space. The physical 
Hamiltonian operator, the energy observable, is to be identified 
among them. (Newtonian) time is its flow/evolution parameter. 
In that sense, we can say that the fundamental symmetry as symmetry 
for the phase space essentially dictates the dynamics of Newtonian 
time, though it does not identify which particular one-parameter 
Hamiltonian flow/evolution is the time flow/evolution. In fact, 
it need not even have the physical Hamiltonian operator as 
representing any element in the Lie algebra. The $H$ generator
in $G(3)$ cannot give the correct time anyway.

To avoid confusion and facilitate the appreciation of some of
the subtle issues presented in our discussions below,  
let us present the structure
and logic of the mathematics and then the physically formulated
theory as an application in general, emphasizing what we see as 
the full picture or the full power of the symmetry formulation.

One can start with an abstract (real) Lie algebra. It is a (real) 
vector space, defined by a set of abstract generators as basis 
vectors, on which a Lie product/bracket is defined by giving the 
Lie bracket between any pair of generators as a linear combination 
of all. The corresponding Lie group can be obtained as its formal 
exponentiation. Another structure of importance to physics is its 
universal enveloping algebra, which is essentially formal complex
polynomials of the generators. One may want to extend that
further to functions or even distributions of the generators which 
can also be seen as elements of the complex group ($C^*$-)algebra
and its extension. Subtle details of the part we need not concerned
much here. We are interested in a representation of all of that, 
and particularly the irreducible ones. Hence, we use the term
symmetry theoretical formulation instead of group theoretical
formulation and use basically the group symbols to denote the
whole mathematical package. Most of the explicit 
analysis focuses on the Lie algebras though. A physical system 
bearing the symmetry would correspond to such a representation 
for which one has a representation (complex vector) space with 
operators representing the generators. The operator 
product matches exactly to the formal product and the commutator 
to the Lie bracket. In the case of quantum mechanics, the 
representation space is the Hilbert space of states, and the 
operator algebra is the algebra of observables. A few elements
in the universal enveloping algebra are of special importance in
the representation theory. These are the Casimir elements. 
They form a set of independent elements which commute with
all the generators hence all other elements. An irreducible
eigenspace of the set of Casimir operators, representing the
Casimir elements, gives an irreducible representation of the
symmetry. The corresponding eigenvalues of Casimir operators,
the Casimir invariants, hence completely characterize the 
representation as an elementary physical system bearing the 
symmetry. We expect the Casimir invariants to be important 
physical properties of the system. Actually, the formulation 
of the physical theory should be considered unsatisfactory 
if the system has a fundamental characteristic that does not 
correspond to a Casimir invariant. With the exception of having 
the Newtonian particle mass as a Casimir invariant, our previous 
formulation a particle of zero spin \cite{070} fully illustrates 
the power of the (relativity) symmetry theoretical formulation. 
Though Ref.\cite{070} still acknowledges the symmetry as 
$\tilde{G}(3)$, the formulation used really only the 
$H_{\!\ssc R}(3)$ part without 
the generator $H$ for time translation. 
We present below the key symmetry issues 
about the theory, focusing on the key notions mentioned above 
but skipping detailed formulation aspects which can be easily 
adapted from Ref.\cite{070}. A serious consideration 
of a composite system, of two particles, is an important part.

Note that most of the presentations of symmetry theoretical 
formulation for the quantum theory are based on the 
Heisenberg-Weyl symmetry $H(3)$ not in a manner to be seen 
as a part of the  $\tilde{G}(3)$ symmetry. In particular, the position 
operators are taken as directly representing generators of the abstract 
Lie algebra, which is really problematic. Such a presentation is, as we 
argue below, incorrect. It also hides the related role of the mass from 
the symmetry picture and goes against the notion of the center of 
mass for a composite system. Some discussions on how the Galilean 
symmetry is realized in the quantum theory, for example Ref.\cite{D}, 
get the part of the results fine.

 In general, the Galilean
symmetry picture is well-presented in many textbooks. But the key issues 
from the formulation perspective are left not clarified. We intend here 
to do exactly that, in the next section. Note that an important paper 
from L\'evy-Leblond \cite{LL} is an exception. It addresses directly 
a symmetry theoretical formulation based on the Galilean symmetry, and 
hence serves as a great background reference for our presentation of the 
formulation. However, there is no explicit discussion in the paper of 
the issues concerning the position operators from the formulation. 
 Situation in Refs.\cite{H,HO} is similarly, though they 
give a nice emphasis on taking the language of the central extension 
and have some interesting discussions on Newtonian mass as the central
charge in relation to the equivalence principle and other things. In 
particular, they still fell short of naming the mass as a Casimir 
invariant. Some aspects of formulation, mostly for the part of spin, 
are presented in the appendix, at least to make the presentation of 
the symmetry theoretical formulation more complete.

Another difference between our presentation and that of Ref.\cite{LL}
 and others in the literature,
is about the time translational symmetry part of the $\tilde{G}(3)$ 
symmetry. Our presentation of the formulation, as given in the
next section, used only the part of the so-called $H_{\!\ssc R}(3)$
of the $\tilde{G}(3)$ without the time translation. We address 
how Newtonian time shows up in the dynamical theory from the
Hamiltonian or symplectic geometry perspective in section ~\ref{sec3}.
The considerations for dropping the time translational symmetry,
hence sticking to use only the $H_{\!\ssc R}(3)$ symmetry in the
formulation are discussed. Let us emphasize here that we are not
proposing that one should drop time translational symmetry from
consideration in physics. The notion of time and the symmetry as in 
the admissible arbitrary choice of the origin of time in a dynamical,
at least so long as we are not looking at cosmology, is no doubt of
fundamental importance. We plead only a case that the symmetry 
should not be included in the fundamental symmetry from which our 
dynamical theory of `nonrelativistic' (quantum) mechanics is to be 
obtained as a representation theory of. We also express 
some further opinion on how time is to be seen from physics in general.
That leads to the last conluding section.

\section{Symmetry Theoretical Formulation of Quantum Mechanics and the  Newtonian Mass}
The Heisenberg-Weyl symmetry $H(3)$ for quantum mechanics
is usually presented in terms of the Lie algebra as 
\bea\label{w}
[X_i, P_j]= i\hbar \delta_{ij} I \;,
\eea
which is a direct abstraction of the Heisenberg commutation 
relation between the position and momentum operators 
$[\hat{X}_i, \hat{P}_j]= i\hbar  \delta_{ij}\hat{I}$. The 
generator $I$ is a central charge, having vanishing Lie brackets 
with all other generators and hence all elements of the seven
generator Lie algebra. In any irreducible unitary representation,
it has to be represented by a real multiple of the identity
operator $\hat{I}$. That real number is then the eigenvalue 
of the operator representation of $I$. In fact, any nonzero 
value gives a unitary representation, the direct integral 
of all is the regular representation \cite{Ta}. Note that the 
Stone-von Neumann theorem really only addresses the case
of an operator picture of the Heisenberg commutation
relation, instead of starting from the above Lie algebra 
for which the abstract generator $I$ cannot be assumed 
to be (represented by) the identity. 
The one-parameter set of irreducible representations, other than
the effectively commutative limit of its vanishing value, is well
appreciated in the mathematics literature \cite{Ta}. While such
an extra nontrivial parameter can be absorbed, it is puzzling if
that is really just a redundancy in the relation between the abstract 
mathematics and physics or if it has other implications. 

Explicitly, one can take for each irreducible representation 
characterized by a $\zeta \ne 0$
\bea&&
{\sqrt{\zeta}} \hat{X}_i (= {\sqrt{\zeta}} x_i) 
= \hat{X}_{i_\zeta}  \quad \leftarrow  \quad  {X}_i \;,
\sea
{\sqrt{\zeta}} \hat{P}_i (= -i\hbar {\sqrt{\zeta}} \frac{\partial}{\partial x^i}) 
= \hat{P}_{i_\zeta}  \quad \leftarrow  \quad  {P}_i \;,
\sea\hspace*{.5in}
{\zeta}\hat{I} =  \hat{I}_{\zeta} \quad \leftarrow \quad  I \;,
\eea
which clearly satisfies $[{\sqrt{\zeta}} \hat{X}_i, {\sqrt{\zeta}}\hat{P}_j]
  = i\hbar \delta_{ij}  {\zeta}\hat{I}$ as the homomorphic  to the 
abstract $[{X}_i,  {P}_j] = i\hbar \delta_{ij} I$ relation as required 
for a representation. That is essentially what is given in the mathematic 
literature \cite{Ta} and adopted in our earlier work \cite{070}. 
We want to emphasize that having the operator representation of the
generators on a definite representation (Hilbert) space, here say of the
Schr\"odinger wavefunctions,  is having a representation for the full 
symmetry set-up. The latter includes the all elements of the 
Lie algebra, Lie group, the universal enveloping algebra, \dots etc. We 
have the full observable algebra with each observable as, say, a function 
of the basic observables of position and momentum.  Together 
with a natural symplectic structure on the Hilbert space and its
projective space, the full dynamical theory can be obtained \cite{070}.
The subtly about the different admissible values of  $\zeta$ 
corresponding to different, mathematically inequivalent, irreducible
representations has otherwise apparently not been noted in the 
literature of physics. An obvious reason is that the presence of $\zeta$ 
has no practical implication so long as a single irreducible representation
is concerned. However, the matter should be look into more carefully.
First of all, the central charge $I$ is effectively a Casimir elements and
its representing operator a Casimir operator.  Or we called them so in 
the sense that they share the same key properties in relation to the
representation theory. Namely,  $I$ has a vanishing Lie bracket 
with all generators, hence representing operator commutes with all 
operators in the theory, giving the eigenvalue $\zeta$ the role of 
a Casimir invariant characterizing an irreducible representation. That 
operator is $ \hat{I}_{\zeta}$, that is simply $\zeta$ times the identity
operator on that particular Hilbert space. One
should wonder why the formulation has such a Casimir invariant
irrelevant to physics, in stark contrast to the usual role of Casimir
invariants, like the mass and spin of the Poincar\'e symmetry.
Another important question is how the $\hat{X}_i$ and $\hat{P}_i$
relate to the picture of Galilean symmetry. In the latter case, one
may appreciate the identification of $\hat{P}_i$ as directly
representing the generators of (spatial) translational symmetry and 
$\hat{K}_i = m\hat{X}_i$ (with $m$ as the Newtonian mass) as 
representing the generators of the Galilean boosts. That actually is 
a question of how to reconcile a symmetry theoretical formulations 
of the quantum theory with $H(3)$ versus $\tilde{G}(3)$ as the
fundamental symmetry. Lastly, but more importantly, the 
apparently harmless $\zeta$ actually causes serious problems
when the formulation is pushed onto composite systems. For example,
one has to confront the question of the composite of two irreducible
representations with different $\zeta$ values.

We have stated that the abstract symmetry of $H(3)$ is part of the
abstract $\tilde{G}(3)$, like the Lie algebra of the former is a subalgebra
of the latter. At the operator level, it looks like there is no problem
between having $[ \hat{X}_i, \hat{P}_j] = i\hbar \delta_{ij} \hat{I}$ and
$[ \hat{K}_i, \hat{P}_j] = i\hbar \delta_{ij} \hat{M}$ with $\hat{K}_i = m\hat{X}_i$
and $ \hat{M} = m \hat{I}$.  The subtly though is that the simple operator
relations, which are physically correct, do not translate well into parallel
results at the abstract Lie algebra level. The usual Galilean picture, with the
Lie algebra for the $H(3)$ seen as
\bea&&
m \hat{X}_i (= m x_i) 
= \hat{K}_{i}  \quad \leftarrow  \quad  {K}_i \;,
\sea
 \hat{P}_i (=  -i\hbar \frac{\partial}{\partial x^i}) 
  \quad \leftarrow  \quad  {P}_i \;,
\sea\hspace*{.5in}
m \hat{I} =  \hat{M} \quad \leftarrow \quad  M \;,
\eea
$[K_i, P_j]=i\hbar \delta_{ij} M$ is the correct one. The usual  $H(3)$ 
picture of $[X_i, P_j]= i\hbar \delta_{ij} I$ is actually incorrect. That gives $m$, 
the eigenvalue of $\hat{M}$ characterizing an irreducible representation for 
the Casimir element $M$, as the Casimir invariant. It is, of course, the 
Newtonian mass. Note that a relation such as $\hat{X}_i = \frac{1}{m}\hat{K}_{i}$ 
does not translate into a fixed relation in the Lie algebra. $m$ is not a scalar in the
Lie algebra. The operator relation is really $\hat{X}_i = {\hat{M}}^{-1} {\hat{K}_{i}}$
as acting on the fixed Hilbert space of the representation. The $m$ value
is different for a different representation, and the inverse $M^{-1}$ is not 
defined in the Lie algebra or even the universal enveloping algebra. One
cannot write $X_i= M^{-1}{\hat{K}_{i}}$ or  $X_i= \frac{1}{m} K_i$. So, 
while $\hat{K}_{i}$, $\hat{P}_{i}$, and $\hat{M}$ are operators directly
representing generators of the Lie algebra, $\hat{X}_i$ do not  directly
represent elements of the abstract structure. In fact, that is exactly what 
it should be, as to be illustrated in an analysis of a composite system, for
example,  of two particles.

  We have for a particle of mass $m_a (>0)$ the
operators $\hat{M}_a=m_a$,  $\hat{K}_{ia}$, $\hat{P}_{ia}$, and 
$\hat{X}_{ia} = \frac{1}{m_a} \hat{K}_{ia}$ acting on a Hilbert space 
${\mathcal{H}}_a$, and similarly for a particle of mass $m_b ( > 0)$ the 
operators $\hat{M}_b=m_b$,  $\hat{K}_{ib}$, $\hat{P}_{ib}$, and 
$\hat{X}_{ib} = \frac{1}{m_b} \hat{K}_{ib}$ acting on a Hilbert space 
${\mathcal{H}}_b$.  The basic point about the symmetry theoretical 
formulation of a composite system is to identify that as the product 
representation of the irreducible representations describing its contributing 
elementary parts. That is actually a straightforward exercise along the 
line of angular momentum addition widely available in quantum mechanics 
textbooks. For the system of two particles, the formulation dictates that 
$\hat{G} = \hat{G}_a \otimes \hat{I} + \hat{I} \otimes \hat{G}_b$, for any
generator or, in fact, any element of the Lie algebra $G$. 
Here, $\hat{G}_a$ and $\hat{G}_b$ are the operators representing $G$ for 
particle $a$ and $b$, respectively; and $\hat{G}$ acts on the Hilbert space 
${\mathcal{H}}_a \otimes {\mathcal{H}}_b$, the direct product of those 
for the individual components.  That is a straight consequence of the 
mathematical logic, which has very important consequences.
For example, we have
\bea
\hat{M} = \hat{M}_a \otimes \hat{I} + \hat{I} \otimes \hat{M}_b \;,
\eea
as well as 
\bea
\hat{P}_i = \hat{P}_{ia} \otimes \hat{I} + \hat{I} \otimes \hat{P}_{ib}  \;.
\eea
The naive expressions give mass and momentum as additive quantities.
In particular, we have a (total) mass $m=m_{\ssc a} + m_{\ssc b}$ for the 
composite system. The three position operators $\hat{X}_i$ however 
cannot satisfy the kind of simple sum relation to those of the individual 
particles. The operators $\hat{X}_{ia} \otimes \hat{I} + \hat{I} \otimes \hat{X}_{ib}$ 
do not bear any physical meaning as position observables. And they do not
have the right Heisenberg commutation relation with the $\hat{P}_i$ above.
That is actually an indicator that $\hat{X}_i$ do not represent generators 
directly. Otherwise, there is no escape from that notion of additive position 
for a composite system. The simple fact has apparently escaped attention 
in earlier discussions. As from the formulation with $K_i$ as generators 
instead, we have
\bea
\hat{X}_i \equiv  \frac{1}{m} \hat{K}_i 
= \frac{1}{m}  (\hat{K}_{ia} \otimes \hat{I} + \hat{I} \otimes \hat{K}_{ib})
= \frac{1}{m} ( m_{\ssc a} \hat{X}_{ia}  \otimes \hat{I} + \hat{I} \otimes m_{\ssc b} \hat{X}_{ib} )\;.
\eea
Hence, we have the notion of the center of mass position operators 
dictated by the  representation theory.

 Many physicists may, understandably, 
be uncomfortable about taking $K_i$ and $M$ in the place of the usually
used $X_i$ and $I$ as generators for the Heisenberg-Weyl symmetry. Most 
simply do not separate the abstract notion of the Lie algebra from its operator 
representation in a physical theory of a particular system, and may not have
paid attention to the exact symmetry treatment of composite systems. Our 
elaboration above gives the mathematical logic clearly. Tracing that logic
without the bias from what has become too familiar, one will be able to agree 
with the author that what we present here is the better way of seeing the 
symmetry structure behind the physics. In fact, it is the only way to avoid 
the unphysical notion of additive position observables. 

The next symmetry to note is the 
rotational symmetry $SO(3)$, given as 
\bea && \label{3}
[J_{ij}, J_{hk}] = i\hbar  (\delta_{jk}J_{ih}+  \delta_{ih}J_{jk}  -\delta_{ik}J_{jh} - \delta_{jh}J_{ik}) \;
\eea
$i,j,\cdot$ goes from 1 to 3 with $J_{ij}=-J_{ji}$. On the Hilbert space for 
quantum mechanics of a spinless particle, the momentum operators give 
translations in $x^i$, and the position operators give translations in $p^i$
 \cite{070}. That, together with the intuitive notion of the operators as 
the position and momentum coordinate observables suggest each set as 
components of a three-vector. That is encoded into the commutation 
relations between them and the familiar picture of the orbital angular
momentum operators $\hat{J}_{ij}  = \hat{L}_{ij} \equiv\hat{X}_i \hat{P}_j - \hat{P}_i \hat{X}_j
=\frac{1}{m} ( \hat{K}_i \hat{P}_j - \hat{P}_i \hat{K}_j )$. At the abstract 
Lie algebra level, we have the combined symmetry of $H_{\!\ssc R}(3)$ as 
a semidirect product (of Lie groups), with the full Lie algebra as
\bea && \label{hr}
[J_{ij}, J_{hk}] = i\hbar  (\delta_{jk}J_{ih}+  \delta_{ih}J_{jk}  -\delta_{ik}J_{jh} - \delta_{jh}J_{ik}) \;
\sea
[J_{ij}, K_k] = i\hbar (\delta_{ik} K_j -\delta_{jk} K_i ) \;,
\qquad
[J_{ij}, P_k] = i\hbar ( \delta_{ik} P_j -\delta_{jk} P_i ) \;,
\sea
[K_i, P_j]= i\hbar \delta_{ij} M \;.
\eea
The ten generator Lie algebra, giving all the nontrivial Lie brackets, is 
a major part of $\tilde{G}(3)$. Only one generator, the `Hamiltonian' 
$H$ is missing. We denote the symmetry by $H_{\!\ssc R}(3)$ and have
presented a detailed picture of the full formulation for quantum theory
for a spinless particle, together with its classical approximation based
on a symmetry contraction in Ref.\cite{070}. Issues related to the 
generator $H$ is the focus of the next section. The $H_{\!\ssc R}(3)$
symmetry has irreducible representations beyond those of $H(3)$. They 
are the ones with nontrivial spin more or less as discussed in Ref.\cite{LL}.
We sketch the part of the story under the current framework in the 
appendix for completeness. 
 
A couple of extra comments are in order.  The important role of mass $m$ 
in the quantization of Newtonian mechanics has been appreciated, in the 
language of projective representation of the Galilean symmetry \cite{AI,LL}. 
Using the latter in the place of the more proper and direct one of the $U(1)$ 
central extension (namely having the central charge $M$ as a generator 
of the Lie algebra), however, does not help in revealing all the important 
features discussed above. Another question is if the representation of 
nonpositive mass is meaningful. The question of $m$ with the `wrong' sign
has been well answered in Ref.\cite{LL}. For technical reasons, it is better
for us to leave the more rigorous statement to the appendix. What is
important is the sign of the Casimir invariant does not matter.
The representations with the same $|m|$ are essentially the same hence one
can assume the mass to be always positive.  Naively, there is no difficulty 
taking things to the $m \to 0$ limit \cite{LL}. However, having zero central 
charge really means the collapse of the Heisenberg commutation relation 
formally. The full implication of the formulation of such a theory may have
to be taken more carefully. In `nonrelativistic' physics, there seems to be no 
room for massless particles anyway.

\section{\boldmath $H_{\!\ssc R}(3)$ versus  $\tilde{G}(3)$ and the 
Dynamical Nature of Time \label{sec3}}

The notion of relativity symmetry as a fundamental symmetry 
of a dynamical theory was mostly brought into physics explicitly 
from Einstein's introduction of Special Relativity. As so, it was
first addressed as the symmetry of the background spacetime,
or in relation to reference frame transformations  of spacetime
admissible in the dynamical theory. The theory itself is to be given
otherwise. For the `nonrelativistic' case at hand, Newton's Laws
dictate the dynamical behavior of particles. A particle 
as an ideal object occupying a definite point in the Newtonian model 
of the physical space at an instant of (the Newtonian) time $t$ is
introduced. It has only a single assumed characteristic, 
namely its mass $m$. Of course, the notions of space and time,
both modeled on Euclidean geometry, and mass as `quantity 
of matter' are abstractions from our intuitions. The particular 
picture of spatial position, time, and mass from the mathematical
models, or even the necessity of such notions, as well as the
relativity symmetry obtained, are `correct' only up to the success 
of the whole theory and within the logic of the particular
formulation. We have seen, for example, substantial 
modifications of the notion of space and time, and in a way
also the notion of mass, in Einstein's theory compared to 
Newton's. We have discussed how quantum mechanics may be 
better looked at as giving a different, quantum, model of the 
physical space \cite{070,078,088}. Here in this article, we present 
what a symmetry theoretical formulation of the theory says about 
mass and time. We have discussed above how the Newtonian 
mass $m$ is really given from the, $H_{\!\ssc R}(3)$ or $H(3)$, 
symmetry as a Casimir invariant for an `elementary' object as 
an irreducible representation -- the particle. That Casimir invariant 
is not there in the classical symmetry. In this section, we look at 
the Newtonian time, and the picture obtained is actually shared 
by the corresponding classical theory in itself. 

The quantum Hilbert space as the representation space for the
particle is its phase space. And it has a symplectic structure, fixed
by the inner product \cite{078,AS}. The Schr\"odinger equation 
is essentially a set of Hamilton's equations of motion. The
Heisenberg equation of motion should be similarly interpreted
with $\frac{1}{i\hbar}[\cdot,\cdot]$ as the exact Poisson bracket, 
the relation of which to the fundamental Lie algebra is obvious.

The basic observables, not including the spin part, are the 
(noncommutative) phase space coordinates $\hat{X}_i$ and 
$\hat{P}_i$. All other observables can be expressed as like
functions of them. There is no time among them. The full theory
is successfully constructed using only the $H_{\!\ssc R}(3)$, even
when nonzero spin is included. Yet, it does offer a notion of 
the Newtonian time $t$ as a real parameter of unitary 
transformations, actually Hamiltonian transformations.
The physical Hamiltonian $\hat{H}_{\mbox{\tiny  phys}}$, or the 
energy observable, is the generating Hamiltonian function. 
Explicitly, we have
\bea
\frac{d}{dt} \left|\phi\rra = \frac{1}{i\hbar}\hat{H}_{\mbox{\tiny phys}} \left|\phi\rra \;,
\qquad
\left|\phi (t)\rra = e^{\frac{t}{i\hbar}\hat{H}_{\mbox{\tiny phys}}} \left|\phi (t=0)\rra \;,
\eea
for any state $ \left|\phi\rra$, or 
\bea
\frac{d}{dt} \hat{A} = \frac{1}{i\hbar} [ \hat{A},\hat{H}_{\mbox{\tiny phys}}] \;,
\qquad
\hat{A}(t)= e^{\frac{-t}{i\hbar}\hat{H}_{\mbox{\tiny phys}}} \hat{A}(t=0) \;
   e^{\frac{t}{i\hbar}\hat{H}_{\mbox{\tiny phys}}} \;,
\eea
for any dynamical variable as observable $\hat{A}$. Not only that
the dynamical equations give a definite notion of the time $t$, this
dynamical time as the evolution parameter from the 
$\hat{H}_{\mbox{\tiny  phys}}$ obviously depends on the interactions
as encoded in the potential part. The correct Newtonian 
time should always correspond to the evolution parameter of any 
dynamical system with whatever admissible, or practical observed, 
interaction in the theory. There cannot be a single fixed 
expression for $\hat{H}_{\mbox{\tiny  phys}}$ in terms of $\hat{X}_i$ 
and $\hat{P}_i$, and certainly no definite element or generator of 
a fundamental symmetry as an extension of the $H_{\!\ssc R}(3)$ that
may correspond to the representation of. In fact, we see no other way 
to talk about time so long as that dynamical theory is concerned. 

The essence of symplectic geometry is that it admits generic 
Hamiltonian flows or transformations of which the dynamical time 
evolution is just one example. In the case of the quantum theory, 
we have any Hermitian operator $\hat{H}_s$ as an element of the 
observable algebra gives mathematically a one-parameter group 
of unitary transformations on the phase space. Under the Schr\"odinger 
picture it satisfies $\frac{d}{ds} =  \frac{1}{i\hbar}\hat{H}_s$. 
On the observable algebra under the Heisenberg picture it satisfies
$\frac{d}{ds} = \frac{1}{i\hbar} [\cdot, \hat{H}_s ]$. What may be the 
physical meaning of the parameter $s$ for a particular $H_s$ is a
different question. $\hat{L}_{ij}$ is, for example, $H_{\theta_{ij}}$
where $\theta_{ij}$ is the angle of the rotation generated. 
$\hat{H}_{\mbox{\tiny  phys}}$ is exactly $\hat{H}_t$. Most of such 
Hamiltonian transformations are not a basic part of the fundamental 
symmetry. That is to say, $\hat{H}_s$ may not represent 
an element of the Lie algebra itself. The formulation does not give 
the Newtonian time any special mathematical position many physicists 
may prefer to have. But that is the exact mathematical logic. Of 
course, the analogous formulation of the `relativistic' theory would 
put Minkowski time on a special footing together with the Minkowski
spatial positions.  The `nonrelativistic' theory as then
an approximation to that would give the Newtonian time as the limit. 

The usual picture of Galilean symmetry $\tilde{G}(3)$, however, has 
a notion of time translational symmetry with the generator $H$ which 
has as the only nontrivial Lie bracket the rest of the,  $H_{\!\ssc R}(3)$,
generators given by
\bea &&
[K_i, H] = i\hbar P_i  \;.
\eea
The full symmetry has an extra Casimir element, $2MH -P_i P^i$, as
well illustrated in Ref.\cite{LL}. With $m$ fixed in a representation,
the extra Casimir invariant is conveniently taken as the real number
${\mathcal V}$ satisfying 
\bea \label{h}
\hat{H} = \frac{1}{2m} \hat{P}_i \hat{P}^i + {\mathcal V} \;.
\eea
Each irreducible representation for $\tilde{G}(3)$ is hence 
characterized by the triple $\{ m,  s, {\mathcal V} \}$. Obviously, 
an irreducible representation as the quantum theory of a particle 
as described above serves as one for the full $\tilde{G}(3)$. Now, 
$\hat{H}$ is obviously not the correct physical Hamiltonian as
\bea
\hat{H}_{\mbox{\tiny  phys}} \equiv \hat{H}_t = \frac{1}{2m} \hat{P}_i \hat{P}^i + V(\hat{X}_i) \;.
\eea 
It admits no potential, {\em i.e.} no nontrivial dynamics. From the 
perspective of generic Hamiltonian transformations in the phase 
space as described above, we have $\hat{H} \equiv \hat{H}_{\tilde{t}}$ 
for some `time' parameter $\tilde{t}$, with, for example, 
$\frac{d}{d\tilde{t}} =  \frac{1}{i\hbar}\hat{H}_{\tilde{t}}
    \equiv \frac{1}{i\hbar}\hat{H}$ acting on a state as in the 
Schr\"odinger equation. In the practical case when the potential
vanished, {\em i.e.} for a free particle, and only in that special and 
dynamically not so interesting case, $\hat{H}$ agrees with $\hat{H}_t$ 
and ${\tilde{t}}$ agrees with $t$.  So from the point of view 
of the dynamical theory, the `time translation'  generated by $\hat{H}$
is only a translation or shift in the value of the $\tilde{t}$ parameter
characterizing points (states) on the curves of constant $\hat{H}$ 
value of the unitary transformation $e^{\frac{\tilde{t}}{i\hbar}\hat{H}}$.
That is actually irrelevant to physics so long as any nontrivial dynamics 
is concerned. The parameter $t$ gives the correct Newtonian time, 
while the parameter $\tilde{t}$ does not.  Moreover, while we
can kind of practically implement the `translation' in the time $t$,
say in resetting the clock, there is no way to implement 
a `translation'  in $\tilde{t}$, unless we can find a way to `tell'
$\tilde{t}$ which has to be based on a system without any 
interaction what-so-ever.

The observable  $\hat{H}$ is not the energy observable as 
$\hat{H}_{\mbox{\tiny  phys}}$, it is essentially only the kinetic energy,
if one insists on having the theory as one of the $\tilde{G}(3)$ symmetry. 
But the quantity ${\mathcal V}$, which is supposed to be a basic 
characteristic of a particle on the same footing as the mass and the spin, 
has never been given any true physical meaning. It is called internal energy 
in Ref.\cite{LL}, which is clearly inaccessible to physical phenomena. It is odd 
to have a Casimir invariant like that, that we have to talk about particles with 
different values of internal energy that has otherwise no relevance to physics.

We believe there is no notion of time in the theory of `nonrelativistic' 
quantum mechanics that goes beyond what that evolution parameter 
$t$ of the physical Hamiltonian as $\hat{H}_t$ can offer. Physical time 
is exactly the time in a dynamical evolution, and there is no way to 
contemplate time without dynamics.  Each of us human beings is 
a dynamic system. Without any dynamics, without physical changes, 
including what happened in our brain, there is no way to appreciate 
a notion of time. We do not see any necessary notion of `time 
translational symmetry' in `nonrelativistic' physics beyond what is 
given by that dynamical Hamiltonian flow either, unless one insists that 
the theory or the fundamental symmetry has to be able to describe the
empty space-time in itself, as a representation\cite{086}. But for what 
purpose. Actually, even our practical physical notion of space, so long 
as a theory of particle dynamics is concerned, can only be the totality 
of all possible position of a particle. That gives the Newtonian space 
from the classical theory. For the quantum theory, it works only all 
possible position is to be taken as all possible eigenstates of the $\hat{X}_i$, 
otherwise, a better picture is offered by the new new notion of 
quantum model of the physical space \cite{078}. There naive notion 
of Newtonian space-time in itself has no physical content and is not 
experimentally accessible. Moreover, we know our `nonrelativistic' 
theory is only an approximation to the `relativistic' theory. The latter 
certainly gives a notion of time as part of the Minkowski spacetime, 
if that can be a comfort for those who do not want to do without 
a notion of Newtonian space-time as a basic part of the symmetry 
theoretical formulation of the dynamical `nonrelativistic' (quantum) 
dynamics. That simply do not work except for trivial dynamics. The
right symmetry should be taken as $H_{\!\ssc R}(3)$ only, not
$\tilde{G}(3)$.

\section{Further Concluding Remarks} 
Applications of full symmetry theoretical formulation, naturally 
coupled with the matching picture of symplectic or Hamiltonian 
dynamics,  of all aspects of the theory for  `nonrelativistic' quantum 
mechanics have been presented in Ref.\cite{070}. The formulation 
includes a symmetry contraction that retrieves all aspects of the 
classical theory as an approximation. The (quantum) relativity 
symmetry used is $H_{\!\ssc R}(3)$ as presented above. We 
present a sketch of the formulation here with mathematically
minor, but physically very important, modifications here. The 
key is to have the naive picture of the abstract Heisenberg-Weyl
symmetry $H(3)$ as having generators $\{X_i, P_i, I\}$ and
operator representation as given in Eq.(2) to be replaced by
generators $\{K_i, P_i, M\}$ and operator representation as 
given in Eq.(3) exactly in line with the Galilean symmetry picture. 
The modification is necessary to avoid having position observables
$\hat{X}_i$ being additive, as like the momentum and mass, in
a composite system. The improved $H_{\!\ssc R}(3)$ symmetry 
theoretical formulation has as Casimir invariants the Newtonian
mass and spin, the former characterizing the different admissible
irreducible representations of the Heisenberg-Weyl symmetry.
The implication is very interesting. It is intriguing that mass as 
a fundamental notion in Newton's theory of particle dynamics 
is to be seen as a remnant of the quantum theory. In fact,
it is dictated by the formulation of the latter. The notion of center 
of mass is also dictated, in relation.

The $H_{\!\ssc R}(3)$ symmetry is short of the usually identified
Galilean symmetry of $G(3)$ or here more properly $\tilde{G}(3)$
by missing the generator $H$ of the `time translation' symmetry. 
Not only that the formulation makes no use of that, but we also 
elaborate on how the dynamical theory as retrieved from the 
symplectic geometry of the representation (Hilbert) space gives 
the notion of the correct Newtonian time. It is the 
parameter of the Hamiltonian transformation/evolution parameter. 
The operator representing the generator $H$ give essentially only the 
kinetic energy and cannot gives the correct time translation in the 
Hamiltonian picture so long as there is nontrivial dynamics. From 
the perspective of the formulation presented, there is no need at 
all to have a notion of Newtonian time or the time translation 
symmetry as a part of the fundamental/relativity symmetry. Of 
course, a picture of spacetime and additive energy-momentum 
can be obtained from the Lorentz covariant `relativistic' 
symmetry theoretical formulation with `nonrelativistic'  
theory as an approximation to that.

In coming to the current perspectives, it is important to take
the mathematics of the symmetry and its representation theory
more seriously than what is done in most of the physics 
literature. Starting with an abstract picture of 
the symmetry, one has to reconsider the relation between its 
basic ingredients and the physical quantities based on which 
we first actually identified the symmetry from. 
 
From the symmetry perspective, there is no reason at all 
to expect the need for anything beyond a single irreducible 
representation to be necessary for the description of an 
elementary system taken, as one that cannot be seen as 
composed of different parts. Composite systems are, in general,
described by reducible representations, as the sum of irreducible 
parts, obtained as the product representation of those for the 
parts. The Casimir invariants are the fundamental characteristics 
of an irreducible representation and the (part of the) system it 
describes. For an elementary system, they are universal for all 
states. For a composite system, the story is more complicated. 
Generally, more than one set of Casimir invariants, each for
an irreducible component, mathematically characterize the
different subspaces of the full phase space of the composite 
system which are invariant under all observables constructed 
from the symmetry. If one starts with a state within one such 
invariant subspace, the description of dynamical features of the 
state certainly need not go beyond that subspace.
 
However, in a quantum theory, there is no trivial answer to 
the question of whether one can prepare an initial state that is 
a linear combination of states in different invariant subspaces. 
An explicit example, commonly discussed in the `relativistic'
setting, is the question of if we can prepare a state as 
a nontrivial linear combination of states of different spins.
The question is particularly interesting in relation to the notion 
of quantum frames of 
reference which has been catching popularity lately (interested
readers are referred to Ref.\cite{19} and references therein).
In particular, a carefully detailed analysis has been presented
in Ref.\cite{L} addressing symmetry issues
with a practical consideration of the relative nature of observed 
quantities, as well as the validity of superselection rules. The key 
concluding statement of ``observable quantities are invariant 
under symmetry and that, in quantum mechanical laboratory 
experiments, the measured statistics pertain not to some 
absolute quantity, but rather to an observable, relative quantity, 
corresponding to the system and apparatus combined, along 
with the appropriate high localisation limit on the side of the 
apparatus" is to be taken as the background to understand our 
discussion of the observables as apparent `absolute quantities'. 
Here, the ``apparatus" can be replaced by, or embodies, the 
physical object as the frame of reference to give physical 
definition to the observables.  The ``high localisation 
limit" essentially corresponds to objects which can be well described 
as classical.  For a composite of two particles with one being 
essentially classical, of a large mass, that conclusion says that 
the $R$-$Q$ variables as relative observables to the center 
of mass describe well the physics of the light, quantum, particle.  
That offers essentially the same picture as given by
the absolute quantity description picture of the 
single-particle system we presented. Though that is what one 
would expect, the solid confirmation of that is as important as 
it is interesting. To truly look at a system of quantum particles, 
however, quantum particles observed from a physical frame 
of reference, the quantum nature of which cannot be neglected, 
more studies, probably along the lines of Refs.\cite{19} and
 \cite{093}, would be needed. 

Finally, we want to emphasize that empty space-time is of little 
interest to physics, the theories of which are to describe physical 
phenomena. Dynamics are geometrically described in the phase 
space. Hence, it should be the symmetry of the latter, rather 
than that of a simple notion of space-time, that is of fundamental 
importance. If relativity symmetry is about the symmetry of
reference frame transformations, it is likewise reference frames 
for the description of dynamics rather than only that of the
space-time that should be relevant. Especially when considering 
the important violations of the naive Newtonian notion of
momentum being the product of mass and velocity,  for 
example in the case of electrodynamics, direct consideration
of reference frames (transformations) for the phase space has
to be taken seriously. The only physical notion of the space is
the totality of all possible positions of a (free) particle, and as 
such can be retrieved as part of the phase space.
Our symmetry theoretical 
formulation of the quantum theory gives at the classical limit, 
the Newtonian single particle phase space as a representation
with the Newtonian space as an irreducible component agreeing
with the coset space picture. In the quantum setting, the Hilbert
space is an irreducible representation, hence cannot be seen
as a product of independent configuration/position space and 
momentum space. No irreducible representation gives exactly
the Newtonian space at the classical limit anyway. That is exactly 
like the fact that the Minkowski 
spacetime as an irreducible representation of the `relativistic' 
symmetry can be definitely split into the Newtonian space and 
the Newtonian time at the `nonrelativistic' limit. 
That further justifies calling the $H_{\!\ssc R}(3)$
(quantum) relativity symmetries and the  single particle quantum 
phase spaces the quantum model of the physical space 
 \cite{070}. That can actually be 
used  to give a picture of quantum mechanics as particle dynamics 
on the quantum/noncommutative geometric model of space
 with the position and momentum observables 
as coordinates \cite{078,088}.

The corresponding picture for the `relativistic' case, with other
important implications, will be addressed in a separate publication \cite{096}.

\appendix*
\section{On Representations of Nonzero Spin}
We focus first on irreducible representations of the $H_{\!\ssc R}(3)$
symmetry with nonzero spin. These are representations that
go beyond those of $H(3)$. They are more or less as presented in 
Ref.\cite{LL} under the $\tilde{G}(3)$ symmetry perspective, there
are subtle but interesting differences between our presentation 
here and that of the latter reference though.  We have the spin 
label $s$ as effectively a Casimir invariant. the actual Casimir 
element is $\frac{1}{2} T_{ij} T^{ij}$, with 
$T_{ij} \equiv M J_{ij} -(K_i P_j - P_i K_j)$, which actually `commute' 
with all generators  (Einstein summation convention assumed, and
generators written with upper $i,j$ indices are the same as the ones 
with lower indices).  Note that $T_{ij}$, like all quadratic Casimir 
elements, do not exist as elements of the Lie algebra. They have to 
be taken as elements of the universal enveloping algebra. Of course 
at the representation level, all the operator sum/product combinations 
are well defined. The Casimir element $\frac{1}{2} T_{ij} T^{ij}$ can 
be seen as sitting in for the $\frac{1}{2} J_{ij} J^{ij}$ of $SO(3)$ which 
gives our familiar angular momentum as a Casimir invariant.

Introducing intrinsic angular momentum operators
$\hat{S}_{ij}\equiv\frac{1}{m} \hat{T}_{ij}$, we have the familiar 
$\hat{J}_{ij} = \hat{L}_{ij} + \hat{S}_{ij}$ for the orbital angular 
momentum operator 
$\hat{L}_{ij}  = \hat{X}_i \hat{P}_j - \hat{P}_i \hat{X}_j$.
As ${T}_{ij}$ commute with all other generators besides the ${J}_{ij}$,  
we can think about a Lie algebra as the direct sum of an $SO(3)$ 
from ${T}_{ij}$ with the invariant subalgebra from the generator set 
$\{K_i, P_i, M, H\}$, the $H(3)$. Irreducible representations of the 
symmetry are then each a simple direct product of irreducible
representations of the two parts. Each such irreducible representation 
serves as an irreducible representation of our $H_{\!\ssc R}(3)$, 
and that exhausts the list. The $SO(3)$, or its double cover, gives the
familiar $(2s+1)$-dimensional representation space of the spin 
components for the intrinsic angular momentum $s(s+1) \hbar^2$ 
as the effective Casimir invariant. It is the 
eigenvalue for the operator
$\frac{1}{2} \hat{S}_{ij} \hat{S}^{ij}$, for integral or half-integral 
values of $s \geq 0$. Each component is a replica of the same  
representation of zero spin as an irreducible representation of the 
$H(3)$ part.  Nonzero spin in  `nonrelativistic' physics is not commonly 
addressed. But spin is really about rotational symmetry, whether 
it is $SO(3)$ or spacetime rotations of $SO(1,3)$, and can be 
addressed independently \cite{LL,LL2} or as approximations to the 
`relativistic' one. Under the essentially $\tilde{G}(3)$ framework, 
an irreducible representation has the labeling characteristics 
$\{ m,  s, {\mathcal V} \}$, where $s$ is the spin label and can 
be mapped, through an anti-unitary transformation, to one of
$\{- m,  \bar{s}, -{\mathcal V} \}$ with $\bar{s}$ denoting the $SO(3)$ 
representation conjugates to that of $s$.  
Hence, the two representations are essentially the same.
The basic picture already presented in Rf.\cite{LL}, of course, 
maintains with only $H_{\!\ssc R}(3)$, only having ${\mathcal V}$ 
dropped from consideration.

Let us further consider a composite system as a product of two
irreducible representations of $\{ m_{\ssc a},   s_{\ssc a} \}$
and $\{ m_{\ssc b},   s_{\ssc b}\}$. As noted, $m_{\ssc a}$ and 
$m_{\ssc b}$ can be taken as positive. The first question is what 
are the admissible values of the Casimir invariants? A closely 
related question is if the representation divides into a number of 
irreducible components. Note that we have, in a simplified form
of the tensor product notation, $\hat{P}_i = \hat{P}_{ia} +  \hat{P}_{ib}$ and  
$\hat{X}_i  = \frac{1}{m} ( m_{\ssc a} \hat{X}_{ia}  + m_{\ssc b} \hat{X}_{ib})$,
and of course also $\hat{J}_{ij} = \hat{J}_{ija} +  \hat{J}_{ijb}$.
To go on with the analysis of the composite system, it is convenient 
to introduce the complementary observables given by the operators 
$\hat{R}_i  \equiv  \hat{X}_{ia}  -  \hat{X}_{ib}$ and 
$\hat{Q}_i  \equiv  \frac{1}{m} (m_{\ssc b}  \hat{P}_{ia}  -  m_{\ssc a} \hat{P}_{ib})$.
The operators all commute with $\hat{X}_i$ and $\hat{P}_i$
while $[\hat{R}_i,  \hat{Q}_j] = i\hbar \delta_{ij}$. For example,
with the Hilbert space tensor product as spanned by simultaneous
momentum eigenstates $\left|{p}_{ia}, {p}_{ib} \rra$, a more 
convenient basis to work with would simply be given by 
$\left|{p}_{i}, {q}_{i} \rra$, simultaneous eigenstates of $\hat{P}_i$
and  $\hat{Q}_i$. $\hat{X}_i$ and $\hat{P}_i$ are the basic dynamic 
observables for the center of mass degrees of freedom, while 
$\hat{R}_i$ and $\hat{Q}_i$ those for the degrees of freedom of 
the relative motion between the particles which is internal to the 
composite system. One can go as far as seeing the two sets as
sets of canonical noncommutative coordinates for the phase
space of the system \cite{078,088}. 

 With a bit of calculation, 
one obtains for the product representation
\bea
\hat{J}_{ij} 
  = ( \hat{X}_i \hat{P}_j - \hat{P}_i \hat{X}_j ) + ( \hat{R}_i \hat{Q}_j - \hat{Q}_i \hat{R}_j ) + \hat{S}_{ija} + \hat{S}_{ijb}  \;,
\eea
giving 
\bea
\hat{S}_{ij} = \hat{R}_i \hat{Q}_j - \hat{Q}_i \hat{R}_j + \hat{S}_{ija} + \hat{S}_{ijb}\;.
\eea
The interpretation is that the relative motion between the 
two particles gives rise to an `orbital' angular momentum which 
is, however, intrinsic to the system as a composite. The latter 
behaves exactly like part of the spin of the composite taken as 
a particle. The `spin' of the full $\hat{S}_{ij}$, {\em i.e.} 
$\frac{1}{2} \hat{S}_{ij}\hat{S}^{ij}$, is effectively a Casimir invariant. 
This `spin' $s$ for the composite has a list of admissible values 
coming from the standard angular momentum addition picture,
 as the sum of the three parts. The different $s$ values correspond 
to different irreducible representations as components of the 
product representation which is reducible. For example, with the 
$\left|{p}_{i}, {q}_{i} \rra$ basis, each set of vectors with fixed 
$q^2= q_i q^i$ span a subspace invariant under the rotations 
generated by $\hat{S}_{ij}$. It is important to emphasize that neither 
$\frac{1}{2} \hat{J}_{ij} \hat{J}^{ij}$ nor $\frac{1}{2} \hat{L}_{ij} \hat{L}^{ij}$,  
$\hat{L}_{ij} \equiv \hat{X}_i \hat{P}_j - \hat{P}_i \hat{X}_j$, 
serves as Casimir invariant, only $\frac{1}{2} \hat{S}_{ij} \hat{S}^{ij}$,
which for a composite system of spin-zero particles has
$\hat{S}_{ij}$ as `orbital' angular momentum 
$\hat{L}_{ij}^{\!s} \equiv \hat{R}_i \hat{Q}_j - \hat{Q}_i \hat{R}_j$, 
does. The $\left|{p}_{i}, {q}_{i} \rra$ basis, even restricted to 
fixed $q^2= q_i q^i$, is however generally not the best basis for 
the description of a particular two-particle system. The irreducible 
component of the product representation has, for the degree 
of freedom for the relative motion, a fixed angular momentum.  
The latter contains both the `orbital' part and the part of the spins 
of the particles. As we appreciate from the physics, a closed
composite system has the total angular momentum, as the
$\hat{S}_{ij}$ here, being conserved, but the interaction between
the parts allows the exchange of angular momentum between
them, including the spin parts with the `orbital' part. Apart from 
the $\left|{p}_{i} \rra$ for the center of mass degree of freedom, 
the degree of freedom for the relative motion is to be 
described in a spin space of finite dimension $2\ell_s +1$, with 
$\ell_s(\ell_s +1) \hbar^2$ being the eigenvalue of the Casimir 
invariant. All that is independent of the actual interaction 
dynamics between the particles. 
 

In relation to the interaction paradigm, the physical Hamiltonian 
or the energy observable for the composite system, neglecting
spin-dependent interactions, is expected to have the form
\bea
\hat{H}_{\mbox{\tiny phys}} = \frac{1}{2m} \hat{P}_i \hat{P}^i + \frac{1}{2\mu} \hat{Q}_i \hat{Q}^i
   + V(\hat{R}_i \hat{R}^i) \;.
\eea
Having $\frac{1}{2} \hat{S}_{ij} \hat{S}^{ij}$ as a Casimir invariant
enforces the potential to be a function of $\hat{R}_i \hat{R}^i$. 
Intuitively, one can see that the interaction potential has to be 
independent of $\hat{X}_i$ and $\hat{P}_i$. In that sense, when one 
writes down a nontrivial dynamical description of a particle, it is really 
the dynamics of $\hat{R}_i$-$\hat{Q}_i$ that one is writing. The
Hamiltonian is invariant under the rotations generated by 
$\hat{J}_{ij}=\hat{L}_{ij}$ and  $\hat{S}_{ij}=\hat{L}_{ij}^{\!s}$. The 
practical Hilbert space is the representation space of an irreducible 
component of the product representation fixed the initial value of 
$\ell_s$, which is spanned by $\left|m_{\ell_s} \rra$, $m_{\ell_s} \hbar$ 
being the eigenvalue of $\hat{L}_{\ssc 12}^{\!s}$.

\bigskip
 {\bf Acknowledgements:}\\
The author thanks H.K. Ting for discussions. He is partially supported 
by research grant numbers  110-2112-M-008-016 and
111-2112-M-008-029
of the MOST of Taiwan.

\end{document}